\documentstyle[12pt]{article}
\def\references{
\section*{{\normalsize \bf References}}
\itemsep 0pt
\list
 {[\arabic{enumi}]}{\settowidth\labelwidth{10}\leftmargin\labelwidth
\labelsep=0pt
 \advance\leftmargin\labelsep
 \itemsep=0pt
 \usecounter{enumi}}
 \def\newblock{\hskip .11em plus .33em minus -.07em}
 \sloppy
 \sfcode`\.=1000\relax \small}

\def\case#1/#2{\textstyle \frac{#1}{#2}}
\begin{document}
% (11pt+2pt)*10lines = 130 pts = 1.8 in
\begin{tabular}{|l|} \hline
To appear in {\it Recent Progress in
Many-Body Theories}, vol. 4, \\
edited by E. Schachinger, {\it et al.}
(Plenum, New York).\\ \hline
\end{tabular}
\vspace*{1.0in}

\noindent
IMPROVED TREATMENT OF FREQUENCY SUMS IN
\newline
PROPAGATOR-RENORMALIZED PERTURBATION THEORIES
\vspace{0.25in}

\noindent
\makebox[1in]{}  J.J. Deisz$^a$, D.W. Hess$^b$, and J.W. Serene$^a$
\newline

\noindent
\makebox[1in]{} $^a$ Department of Physics
 \newline
\makebox[1in]{}      Georgetown University
 \newline
\makebox[1in]{}      Washington, D.C. \ \ 20057-0995
 \newline

\noindent
\makebox[1in]{} $^b$ Complex Systems Theory Branch
 \newline
\makebox[1in]{} Naval Research Laboratory
 \newline
\makebox[1in]{} Washington, D.C.  \ 20375-5000
\parindent 0.15in
\parsep 50pt
\vspace{-0.15in}
\newline
\parskip=13pt
\parindent=0.3in

{\flushleft{\bf Introduction}}

\vspace{0.05in}

\noindent
This work stems from calculations
for Hubbard \cite{shmiami,shrapid,shmb} and Anderson
lattice \cite{alprl, alprb} models in a self-consistent conserving
Green's function scheme
\cite{luttinger_ward,baym_phi}
known as the the fluctuation exchange
approximation (FEA) \cite{bickers_fea}.
For the 2D Hubbard model,
special features of band structure, such as Fermi surface nesting
\cite{virosztek} and van Hove singularities
near the Fermi surface \cite{lee_vanhove,kane_vanhove}, lead to
anomalous frequency and momentum dependences of the self-consistent
self-energy \cite{shmb,hess_jpcs}.
At half filling the FEA self-energy develops a frequency dependence
similar to that proposed for a
marginal Fermi liquid \cite{varma_mfl}, and the
spin-fluctuation propagator appears to move exponentially close
to an instability with decreasing temperature.  When
the spin-fluctuation propagator is sufficiently close to this instability,
we have been unable to obtain stable converged solutions.
For the half-filled 3D Hubbard model, where an antiferromagnetic phase
transition is expected at finite temperature, we have studied
the fully self-consistent spin response to a staggered magnetic
field \cite{deisz}.
The results are qualitatively similar to those in 2D \cite{bickers_al},
and show no magnetic
order for a range of
$U$ and $T$ well within the antiferromagnetic phase expected from
Quantum Monte Carlo simulations \cite{scalettar}.

For the Anderson lattice model, we have observed the
evolution of a coherent quasiparticle state with decreasing
temperature along with a substantial enhancement in the effective
mass \cite{alprl}.  As in the Hubbard model, for parameters
relevant to  correlated electronic systems,
the spin fluctuation propagator is very near an instability
at low temperatures. We would like to calculate reliably the
entropy and specific heat as a function of temperature, to
elucidate the apparent transformation from a lattice of
local moments in a sea of `ordinary' conduction electrons
to a band of highly renormalized quasiparticles.

The only numerical approximation (other than finite machine precision)
in our previous implementations of the FEA has been the treatment of
the high-frequency
tails of Green's functions, self-energies, etc.  Although we are primarily
interested in understanding the low-energy excitations, high-energy
processes make important contributions to effective masses,
susceptibilities, total energies, etc., and numerical approximations
in treating these processes must be controlled, and
so far as possible eliminated, to obtain reliable results for the problems
described above.
In this paper we describe a new approach to handling high-frequency tails.
We decompose the single particle
Green's function into two parts,
\begin{equation}
G({\bf k}, \varepsilon_n) = g({\bf k}, \varepsilon_n)
   + \tilde{G}({\bf k}, \varepsilon_n),
\end{equation}
where, as we will make more precise below,
$\tilde{G}$ contains only `low'
frequency parts of $G$ and $g$ contains the leading `high' frequency parts
(through some order).    The important observation is that relatively
little information is contained in the high frequency tails of the
Green's function. The crucial trick is to find
an analytic form for $g$ that describes the high frequency
behavior of $G$ accurately and leads to tractable analytic
expressions for the contributions from $g$ to
susceptibility bubbles, $T$-matrices and self-energies.
Most of the detailed information about correlations resides
in $\tilde{G}$, which is much less sensitive to the frequency cutoff than
was the original $G$. In this language, most
previous approaches to solving the FEA numerically correspond
to taking $g$ to be identically zero \cite{pao_rg}.

Taking advantage of massively parallel computers requires
scalable algorithms that perform
efficiently for a wide range of problem sizes using
virtually any number of processors.
To this end, our algorithm solves the equations of the
FEA iteratively, making use of discrete Fourier transforms
at various stages of the calculation to make each step
embarrassingly parallel.  To motivate our new
approach, we first sketch a less accurate but more straightforward
way to calculate the FEA self-energy.

{\flushleft{\bf Standard Implementation of the
Fluctuation Exchange Approximation}}

\vspace{0.05in}

\noindent
Central to propagator renormalized perturbation theory
is Dyson's equation, relating the renormalized
propagator $G$ to the self-energy $\Sigma$,
\begin{equation}
G^{-1}({\bf k}, \varepsilon_n) =
           G_0^{-1}({\bf k}, \varepsilon_n) -
           \Sigma({\bf k}, \varepsilon_n),
\label{dyson}
\end{equation}
where $G_0$ is the Green's function of the non-interacting system.
For simplicity we will discuss only the simplest paramagnetic Hubbard
model, and will include in the self-energy
only the proper second-order diagram and the contribution from
exchanged spin fluctuations; including the density and pairing fluctuations
is straightforward and involves no new matters of principle.
The fluctuation propagator is constructed from a
susceptibility bubble which is a
convolution product of renormalized propagators,
\begin{equation}
 \chi_{ph} ({\bf q}, \omega_m)  =  - {{T} \over {N}} \sum_{{\bf k}, n}
  G({\bf k + q}, \varepsilon_n + \omega_m) G({\bf k}, \varepsilon_n).
\label{bubble_convolution}
\end{equation}
In terms of $\chi_{ph}$ the fluctuation propagator ($T$-matrix) is simply
\begin{equation}
   T ({\bf q}, \omega_m) =  \frac {3}{2} \ \left[
    \frac{U \chi_{ph}({\bf q}, \omega_m)^2}{1 - U \chi_{ph}({\bf q},
         \omega_m)} \right],
\label{tmatrix}
\end{equation}
and the self-energy is a convolution of the Green's function
with the sum of the $T-$matrix and the susceptibility bubble,
\begin{equation}
\Sigma({\bf k}, \varepsilon_n)  = U^2 {{T} \over {N}}
\sum_{{\bf q}, \omega_m}
 G({\bf q + k}, \omega_m + \varepsilon_n) [ \chi_{ph} ({\bf q}, \omega_m) +
   T({\bf q}, \omega_m) ] .
\label{sigma_ph_convolution}
\end{equation}

\vspace*{-0.1in}
To obtain the self-consistent self-energy, one starts with
a guess for $G$ ({\em e.g.} $G_0$) and calculates the self-energy from
Eqs.(\ref{bubble_convolution}-\ref{sigma_ph_convolution}).
The resulting $\Sigma$ is then used
in Dyson's equation to update the propagator, and this procedure is
iterated to some level of approximate self-consistency.

The sums in these equations extend over all momenta and all frequencies,
but in a numerical calculation we can include only
a finite number of terms.
For the momentum sums this is simply equivalent to taking a finite
lattice with periodic boundary conditions.
Truncating the frequency sums admits no simple physical interpretation,
however, and providing an alternative to truncation is the focus of
this paper.  From the perspective of this section, the most straightforward
procedure is to introduce a sharp cutoff, by setting
the Green's function, self-energy, etc. equal to zero for all
$| \varepsilon_n | > \varepsilon_c$,
which we will call the {\it sharp cut-off scheme}.
This leads to highly artificial behavior of the susceptibility,
$T$-matrix, and self-energy with increasing frequency.
More important from the point of view of instabilities and phase
transitions, the self-energies and susceptibilities at low frequencies
lose high-frequency contributions from Green's functions,
susceptibilities, and T-matrices.

{\flushleft{\bf Posing the Problem Another Way}}

\vspace{0.05in}

\noindent
The fluctuation-exchange approximation for the Hubbard model (and related
lattice models such as the Anderson lattice model) has a special
feature that makes it especially well-suited for a fine-grained SIMD
parallel computer such as the Connection Machine:
the bare interaction is completely local in space and time, and
the approximation does not introduce any
nonlocal effective
interactions.  As a result,
all equations of the theory can be solved completely in parallel at each point
in either (${\bf k}, \varepsilon_n$) space (Dyson's equation and the
$T$-matrix equations) or
(${\bf r}, \tau$) space (susceptibilities and self-energy), without the need
to evaluate directly any convolutions.
In the (${\bf r}, \tau$) representation,
the susceptibility bubble is simply
\begin{equation}
  \chi_{ph}({\bf r}, \tau) = - \; G({\bf r}, \tau) G(- {\bf r}, - \tau),
\end{equation}
and the self-energy is
\begin{equation}
  \Sigma({\bf r}, \tau) = U^2[ \; \chi_{ph}({\bf r}, \tau)
      + T_{ph}({\bf r}, \tau)] G({\bf r}, \tau) ,
\end{equation}
while the natural representations of the  T-matrix and Dyson equation
are given by Eqs.\ (\ref{dyson}) and (\ref{tmatrix}) above.
{}From this point of view, a possible approach is to begin from
$G_0(\varepsilon_n)$
with a sharp high-frequency cutoff as before, and to transform back and
forth between (${\bf k}, \varepsilon_n$) and (${\bf r}, \tau$)
using fast Fourier
transforms (FFTs);
we will call this the $\varepsilon_n$-{\it scheme}.
This yields Green's functions at a discrete set of
evenly spaced $\tau$-points between 0 and $\beta$, but the sharp cutoff
causes endpoint ringing near $\tau = 0$ and
$\tau = \beta$.  Because many physical quantities come from the Green's
function at precisely these endpoints, a better approach is to begin the the
calculation from the exact $G_0(\tau)$ sampled on a uniform mesh of
$\tau$-points, which, after an FFT, has the effect of introducing
a gentle high-frequency cutoff in $G_0(\varepsilon_n)$.  The previous
calculations described in the introduction use this approach,
which we will call the $\tau$-{\it scheme} \cite{shmb}.

{\flushleft{\bf
Contributions from High-Frequency Parts of the Green's Function}}
\vspace{0.05in}

\noindent
Repeated integration by parts of the Fourier integral
for $G({\bf r},\varepsilon_n)$
shows that the discontinuities of the Green's function and its
derivatives at $\tau=0$ determine the high-frequency behavior
of the Green's function,
\begin{equation}
\label{asymptotic}
G({\bf r}, \varepsilon_n) =
{ {-\Delta G({\bf r})}\over{i\varepsilon_n}}
+ \; \cdot\cdot\cdot\cdot  \;
+ {{(-1)^{p+1}\Delta G^{(p)}({\bf r})}\over{(i\varepsilon_n)^{p+1}}}
+{{(-1)^{p+1}}\over{(i\varepsilon_n)^{p+1}}}
\int^{\beta}_0 e^{i \varepsilon_n \tau} {{\partial^{p+1}G({\bf r},\tau)}
\over{\partial \tau^{p+1}}}\,d\tau,
\end{equation}
where
\begin{equation}
\label{discontinuity}
\Delta G^{(p)}({\bf r}) \equiv {{\partial^p G({\bf r},\tau)}\over
{\partial \tau^p}}\Bigl|_{\tau = 0^+} - \,{{\partial^p G({\bf r},\tau)}\over
{\partial \tau^p}}\Bigl|_{\tau = 0^-}.
\end{equation}
Substituting this expression for the Green's functions and a
 similar high-frequency expansion for $\Sigma$ into
Dyson's equation leads to expressions
for the discontinuities of the renormalized
propagator in terms of unrenormalized single-particle energies
$\xi_{\bf k}$ and discontinuities in $\Sigma$,
\begin{equation}
\label{dyson_asy}
G({\bf k},\varepsilon_n) \sim {{1}\over{i\varepsilon_n}} + {{\xi_{\bf k} +
\Sigma_H}\over{(i\varepsilon_n)^2}} +
{{(\xi_{\bf k} +
\Sigma_H)^2 - \Delta \Sigma({\bf k})}\over{(i\varepsilon_n)^3}}
+ \; \; \cdot\;\cdot\;\cdot
\end{equation}
where $\Sigma_H$ is the Hartree-Fock contribution to the self-energy.
For the Hubbard model with nearest-neighbor hopping only,
Eq.\ (\ref{dyson_asy})
gives the first two discontinuities as
\begin{equation}
\Delta G({\bf r}) = - \delta_{{\bf r}, {\bf 0}},\;\;\;\;
\Delta G^{\prime}({\bf r}) = (- \mu + U n/2)\, \delta_{{\bf r},
{\bf 0}}
- t \,\delta_{|{\bf r}|, 1},
\label{delta_g}
\end{equation}
where $\mu$ is the chemical potential, $t$ is the hopping matrix element,
$n$ is the (self-consistent) density and $U$ is the Hubbard
interaction.

We write $G$ as the sum of an analytic part $g({\bf r}, \varepsilon_n)$
containing the leading high-frequency behavior, and a part
$\tilde{G}({\bf r}, \varepsilon_n)$ represented numerically
up to a maximum frequency $\varepsilon_c$,
\begin{equation}
G({\bf r},\varepsilon_n) =
\tilde{G}({\bf r},\varepsilon_n) + g({\bf r},\varepsilon_n).
\label{green_split}
\end{equation}
A simple analytic form for $g({\bf r},\varepsilon_n)$ that
includes the discontinuities of Eq.\ (\ref{delta_g}) is
\begin{equation}
g({\bf r},\varepsilon_n) = -\Delta G({\bf r})\, Q_0(\varepsilon_n, x_0({\bf
r}))
+ \Delta G^{\prime}({\bf r})\, Q_1(\varepsilon_n, x_1({\bf r}))
\label{g_analytic}
\end{equation}
with
\begin{equation}
Q_0(\varepsilon_n, \,x_0({\bf r})) =
    {{1}\over{2}} \; \biggl[
 {{1}\over{i\varepsilon_n - x_0({\bf r})}}
                   + {{1}\over{i\varepsilon_n + x_0({\bf r})}}
                     \biggr]
\;\;\;  \to  {{1}\over{i\varepsilon_n}} \;\;\; \hbox{for}\;
\varepsilon_n \to \infty,
\label{q0}
\end{equation}
\begin{equation}
Q_1(\varepsilon_n, \,x_1({\bf r})) =
{{-1}\over{2 x_1({\bf r})}}\;\biggl[
{{1}\over{i\varepsilon_n + x_1({\bf r})}}
- {{1}\over{i\varepsilon_n - x_1({\bf r})})}
\biggr]
\;\;\; \to {{1}\over{(i\varepsilon_n)^2}}
\;\;\; \hbox{for}\;
   \varepsilon_n \to \infty.
\label{q1}
\end{equation}
The discontinuity and derivative discontinuity
of $G$ are included in $g$
independent of the
choices for $x_0({\bf r})$ and $x_1({\bf r})$.
We choose these parameters by setting
$\tilde{G}({\bf r},
0)$ for $|{\bf r}| = 0,\,1$ and $\tilde{G}^{\prime}( {\bf 0},0)$
equal to zero; we show below that this choice is optimal when
forming the second-order self-energy.

\begin{figure}[htb]
\vspace{2.325in}
\includegraphics{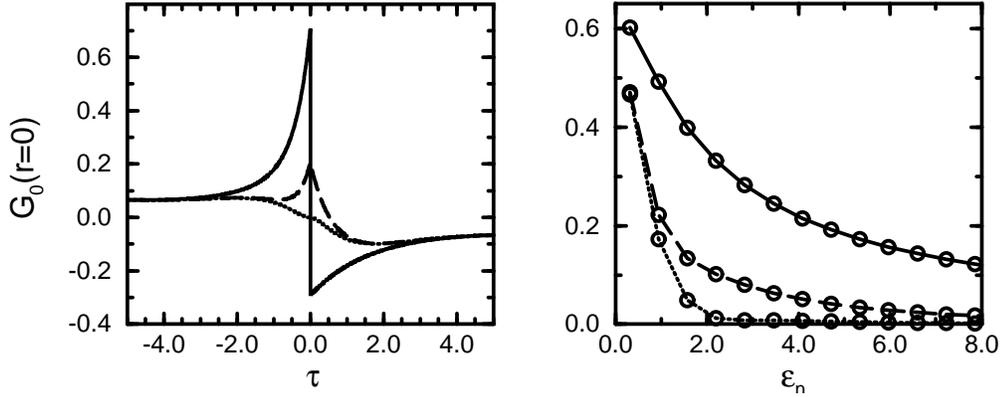}
\center{
\parbox{5.5in}{
{\small
\caption{
The non-interacting
Green's function  $G_0$ at $r = 0$ as a function of $\tau$ (left)
and its modulus as a function of $\varepsilon_n$ (right) for a
1D Hubbard model for $T=0.1t$ and $\mu=1.2t$.
The solid curve is $G_0(r= 0)$, the dashed curve
is $G_0(r=0) - Q_0$, and the
dotted line is $\tilde{G}(r= 0) = G_0(r= 0) - Q_0 + \mu\,Q_1$.
The removal of discontinuities at $\tau=0$ in the part
of $G_0$ represented numerically
corresponds to removing high-frequency tails in
$\varepsilon_n$-space; $\tilde{G}_0(\varepsilon_n)$ has nearly vanished for
frequencies greater than a bandwidth $W = 4t$. Note that all energies
are measured in units of the hopping matrix element $t$.
}
\label{g_r1}
}
}}
\end{figure}

In Fig. (\ref{g_r1}) we illustrate
this decomposition of the non-interacting
Green's
function
in both $\tau$ and $\varepsilon_n$ space.
The solid curve is the full $G_0$, the dashed line is $G_0 - Q_0$, and
and the dotted line shows the final numerical
part, $\tilde{G}_0= G_0 - Q_0 + \mu\,Q_1$,
which is
represented on a discrete $\tau$-mesh and transformed
with an FFT.  The smoother and smaller $\tilde{G}_0$ has a spectral
weight that is effectively confined to low frequencies so that
the errors introduced by $\varepsilon_c$ are
much smaller than those incurred in an FFT of the full Green's
function in any of the standard cutoff schemes.
Analytic terms are Fourier transformed
exactly and functional forms keep track of contributions
to infinite frequency.

As described earlier,
the FFT method takes advantage of the relative simplicity
of expressions
like that for the second-order self-energy in $({\bf r}, \tau)$ space,
\begin{equation}
\Sigma_2({\bf r},\tau) = - U^2 G({\bf r}, \tau) \,G({\bf r}, \tau)
\,G(-{\bf r}, -\tau).
\end{equation}
Using our decomposition of $G$,
a part of this expression can be calculated analytically,
\begin{equation}
\sigma_2({\bf r}, \tau) = - U^2 g({\bf r}, \tau) \,
g({\bf r}, \tau)\, g({\bf -r}, -\tau)
\end{equation}
since this consists of simple functions
of $\tau$ with analytic Fourier transforms.
We choose $x_0({\bf r})$ and $x_1({\bf r})$
so that $\sigma_2({\bf r}, \tau)$ contains
the leading discontinuities of $\Sigma({\bf r},\tau)$. This choice
is optimal in that the remaining numerical piece is continuous to
second order at $\tau = 0$.  For example, the leading discontinuity in
$\Sigma_2({\bf r}, \tau)$ is given by
\begin{eqnarray}
\Delta \Sigma_2({\bf r}) & \equiv &
\Sigma_2({\bf r}, 0^+)
- \Sigma_2({\bf r}, 0^-) \\ \nonumber
                         & = & - U^2 \,
G({\bf 0}, 0^-)\, G({\bf 0}, 0^{+}) \,\Delta G({\bf r}),
\end{eqnarray}
which is finite only for ${\bf r}=  {\bf 0}$.
Since $x_0$ and $x_1$ are chosen
such that $\tilde{G}({\bf 0}, 0)$ vanishes, this
discontinuity is contained entirely in $\sigma_2({\bf r},\tau)$,
\begin{equation}
\Delta \Sigma_2({\bf r}) =
-U^2 \,
g({\bf 0}, 0^-)\, g({\bf 0}, 0^{+})\,\Delta g({\bf r})
\equiv \Delta \sigma({\bf r}).
\end{equation}
In Fig.\ ({\ref{sig_sub}), we show
$\Sigma_2$ calculated with $G = G_0$
as a solid curve and the numerical
part, $\Sigma_2 - \sigma_2$, with
a dotted curve.
It is only this numerical function
whose Fourier transform is approximated with an FFT.
As shown in
Fig.\ (\ref{compare})
this leads to a large reduction in
error in the self-energy
in $({\bf k}, \varepsilon_n)$ space
at all frequencies
in comparison to the errors found with the sharp cut-off,
$\tau$--, and $\varepsilon_n$-- schemes.

\begin{figure}[htb]
\vspace{2.439in}
\includegraphics{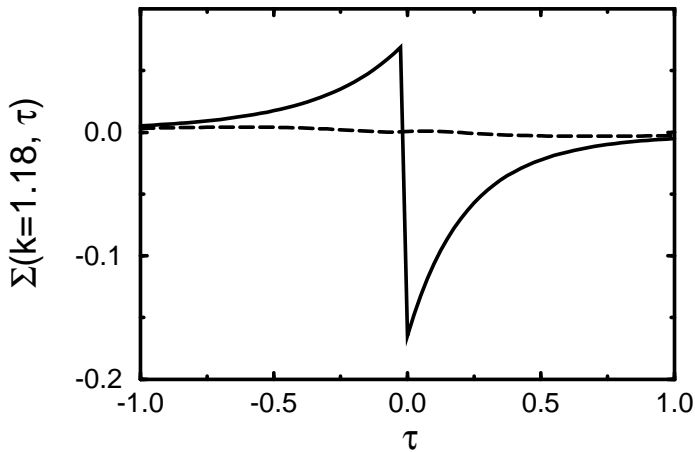}
\center{\parbox{5.5in}{
{\small
\caption{ Full second-order self-energy (solid) calculated
with the bare propagator $G_0$ for a 1D Hubbard model
with $U=t$, $\mu = -0.7t$ and $T=0.04t$ and the
the numerical part (dashed) obtained by subtracting an
analytic contribution evaluated using $g$.
}
\label{sig_sub}
}
}}
\end{figure}

\begin{figure}[htb]
\vspace{2.425in}
\includegraphics{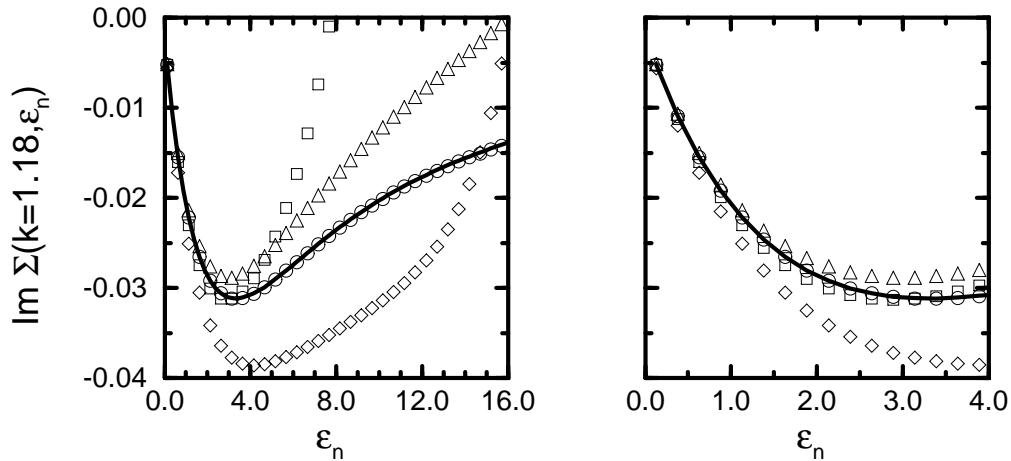}
\center{\parbox{5.5in}{
{\small
\caption{ The second-order self-energy calculated using $G_0$
for the 1D Hubbard model with $U=t$, $\mu=1.4$, and $T=0.04$.
The exact result (solid line), is compared with
this method $(\circ)$, the sharp cut-off method
$(\Box)$, the $\tau$-scheme $(\triangle)$, and
the $\varepsilon_n$-scheme $(\diamond)$. For all frequencies
this method leads to significantly more accurate results
than the traditional frequency cut-off schemes.
Note that every other point is plotted in
the left panel for greater clarity.
 }
\label{compare}
}
}}
\end{figure}

In a self-consistent calculation, the optimal parameters
are adjusted iteration by iteration as correlations
change the $\tau \to 0$ values of the Green's function
and its derivatives.
In Fig. (\ref{self_consistent}) we show the $\varepsilon_0$ point
of the self-consistent second-order self-energy
as a function of the number of points kept in the
representation for $\tilde{G}$
compared to that obtained with the $\tau$-scheme.
As this figure illustrates, the improvements realized
in Fig. (\ref{compare}) are also realized in a self-consistent
calculation.
In particular, as shown for the self-energy at its
lowest frequency, $\varepsilon_0 = \pi T$,
substantially fewer points are
required for this method to achieve the infinite frequency cut-off
limit (obtained with, say, the $\tau$-scheme) to acceptable accuracy.

\begin{figure}[htb]
\vspace{2.5in}
\includegraphics{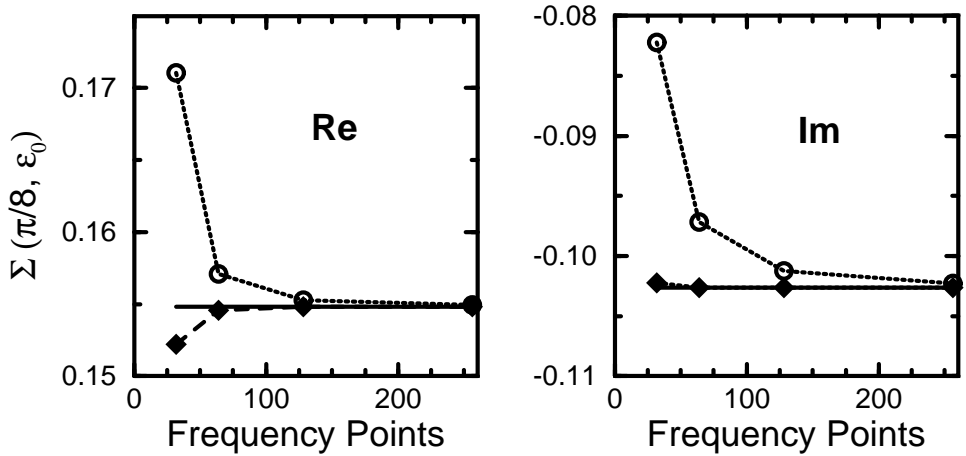}
\center{\parbox{5.5in}{
{\small
\caption{The self-consistent second-order
self-energy at $\varepsilon_0$ for the 1D Hubbard model with $U=4t$,
$T=0.1$ and $\mu=-0.5$ as a function of the number of frequency
points kept in the numerical part of $G$. The rate of convergence
with respect to the number of frequency points obtained with
this method (filled diamonds) is substantially improved with
respect to the ordinary $\tau$-scheme (open circles).  The solid
line represents the $\varepsilon_c \rightarrow \infty$ limit of
the ordinary $\tau$-scheme.
 }
\label{self_consistent}
}
 }}
\end{figure}

It is possible to achieve results with higher accuracy
and more rapid convergence by adding more
analytic terms in $g({\bf r},\tau)$ to include
higher-order derivative discontinuities at $\tau = 0$.
The optimal parameter choices for each function
are again determined by requiring that
the leading discontinuities in products such as
the second-order self-energy are contained entirely
in analytic terms. This requirement is satisfied if
the $x_i({\bf r})$ are chosen so
that the discontinuities and the values of
$G({\bf r}, \tau)$, $G^{\prime}({\bf r},\tau)$, etc.
at $\tau = 0$ are contained in analytic terms
to the fullest extent possible with the available
parameters.

For the full fluctuation exchange approximation, we introduce
analogous asymptotic expansions for the particle-hole
and particle-particle $T$-matrices, and again
represent analytically the leading asymptotic
behavior.
The discontinuities of the $T$-matrix are
determined by the discontinuities of the particle-hole
and particle-particle bubbles, which are
in turn
determined by the single-particle Green's functions.
The optimal parameters for the analytic part of
the $T$-matrix are
again chosen so that
the leading discontinuities in the
self-energy terms $\Sigma({\bf r},\tau) = T({\bf r},\tau)
G({\bf r},\tau)$ are contained in the expressions which
can be
treated analytically, $\sigma({\bf r}, \tau) =
t({\bf r}, \tau) g({\bf r}, \tau)$.
This requirement is satisfied if the values of
$T({\bf 0}, 0)$,
$T^{\prime}({\bf 0}, 0)$, etc.
are contained entirely in the analytic part of the
t-matrix, $t({\bf r},\tau)$.

{\flushleft{\bf Calculation of Thermodynamic Quantities}}

\vspace{0.05in}

\begin{figure}[htb]
\vspace{2.52in}
\includegraphics{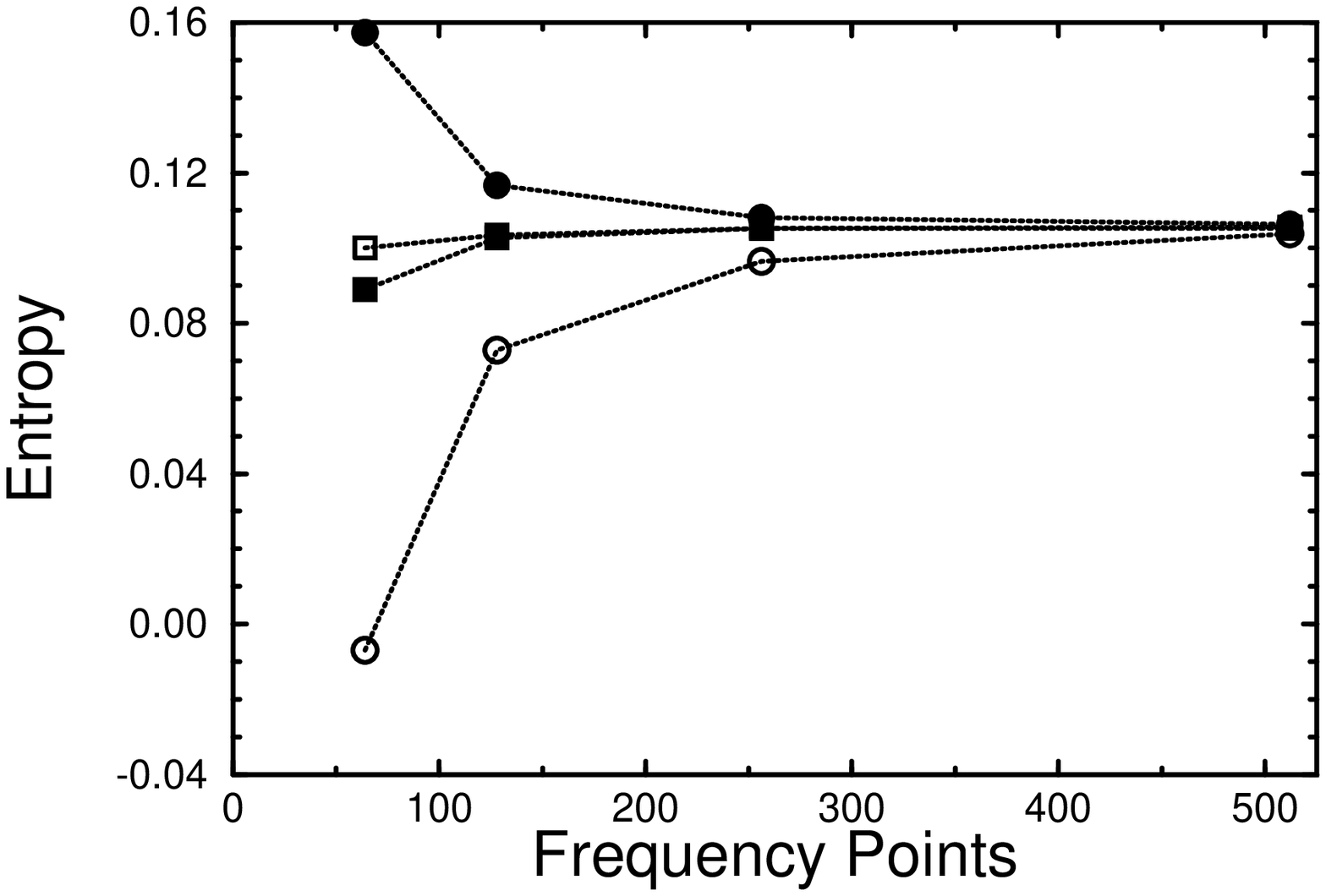}
\center{\parbox{5.5in}{
{\small
\caption{ Self-consistent calculation for
the entropy for the 3D Hubbard model
with $16^3$ sites, $T=0.1$, $n = 0.5$, and $U=4$ as a
function of the number of frequency points.
The open symbols are obtained using the formula
$S = - \partial F(T,N)/\partial T |_N$ and the closed symbols
from the formula $
S = (E - F)/T$.  The results for this method (squares) converge
much more rapidly than those from the $\tau$-scheme (circles).
}
\label{entropy}
}
}}
\end{figure}

\noindent
Thermodynamic properties obtained from the
grand thermodynamic potential calculated in a conserving
approximation (such as the FEA) are guaranteed to be consistent
with those obtained from a direct calculation involving the
self-consistent Green's function and self-energy \cite{luttinger_ward}.
In general, thermodynamic properties and the grand thermodynamic potential
are particularly sensitive to the high-frequency behavior of
the propagator.  A familiar example of the sensitivity  to
high-frequency parts of the Green's function is the slowly
converging frequency sum that results when
the density is calculated by tracing the Green's function.
It is not surprising that thermodynamic properties
have proven difficult to calculate since changes in
temperature produce only small
relative changes in quantities like the free energy, which may
be smaller than a fictitious temperature dependence
introduced by the handling of the high frequency cut-off.
In Fig.\ (\ref{entropy}) we show that the calculation
of the entropy can be achieved keeping a modest
number of Matsubara frequencies when the method presented
in this communication is used.
Here, the entropy $S$ is computed two ways: (1) by numerically
evaluating $S = - \partial F(T,N)/\partial T|_N$ (open symbols)
where $F$ is the Helmholtz free-energy, and (2) by
evaluating $S = (E - F)/T$ where $E$ is the total energy
(closed symbols).
Again, this method (squares) produces more accurate
results for a given number of frequency points than
the $\tau$-scheme (as shown in the figure, the latter can
even yield unphysical values of the entropy when a small
number of frequency points are kept in frequency sums).
Reliable thermodynamic calculations based on self-consistent
perturbation theories may prove helpful in understanding
the thermal properties of interacting quantum systems,
especially since the small system sizes typical of
exact methods make accurate calculations of thermodynamic
properties problematic.

%\newpage

{\flushleft{ \bf Acknowledgments}}

\vspace{0.05in}

\noindent
We thank A.Y. Liu for a critical reading of this manuscript.  This work
was supported in part by a grant of computer time from the DoD HPC
Shared Resource Center, Naval Research Laboratory Connection Machine
facility CM-5 / CM-200.

\end{document}